# The SWAP EUV Imaging Telescope.
# Part II: In-flight Performance and Calibration

J.-P. Halain[1] · D. Berghmans[2] · D. B. Seaton[2] · B. Nicula[2] · A. De Groof[2,3] · M. Mierla[2,4,5] · A. Mazzoli[1] · J.-M. Defise[6] · P. Rochus[1]

**Abstract**. The Sun Watcher with Active Pixel System detector and Image Processing (SWAP) telescope was launched on 2 November 2009 onboard the ESA PROBA2 technological mission and has acquired images of the solar corona every one – two minutes for more than two years. The most important technological developments included in SWAP are a radiation-resistant CMOS-APS detector and a novel onboard data-prioritization scheme. Although such detectors have been used previously in space, they have never been used for long-term scientific observations on orbit. Thus SWAP requires a careful calibration to guarantee the science return of the instrument. Since launch we have regularly monitored the evolution of SWAP's detector response in-flight to characterize both its performance and degradation over the course of the mission. These measurements are also used to reduce detector noise in calibrated images (by subtracting dark-current). Since accurate measurements of detector dark-current require large telescope off-points, we have also monitored straylight levels in the instrument to ensure that these calibration measurements are not contaminated by residual signal from the Sun. Here we present the results of these tests, and examine the variation of instrumental response and noise as a function of both time and temperature throughout the mission.

## 1. Introduction

The *Sun Watcher with Active Pixel System detector and Image Processing* (SWAP), which is part of the PROBA2 payload, is a compact instrument that continuously observes the solar corona in the extreme ultraviolet (EUV) in a narrow bandpass with peak at 17.4 nm (Defise *et al*., 2007 and Seaton *et al*., 2012).

[1] Centre Spatial de Liège, Université de Liège, avenue Pré Aily, 4031 Angleur, Belgium, jphalain@ulg.ac.be

[2] Royal Observatory of Belgium, Avenue Circulaire 3, 1180 Brussels, Belgium

[3] European Space Agency, Department of Science and Robotic Exploration, Noordwijk, Netherlands

[4] Institute of Geodynamics of the Romanian Academy, Jean-Louis Calderon 19-21, Bucharest-37, Romania

[5] Research Center for Atomic Physics and Astrophysics, Faculty of Physics, University of Bucharest, Romania

[6] Institut d'Astrophysique et de Géophysique, Université de Liège, 4000 Liège, Belgium

In addition to its scientific mission, the instrument was also built to demonstrate the usability of complementary metal-oxide-semiconductor active-pixel sensor (CMOS-APS) technology for long-term, in-space scientific applications.

The SWAP sensor is based on the High Accuracy Star-tracker (HAS). It is a 1k x 1k device, of 18 μm pixel pitch, CMOS-APS front-side-illuminated device that is sensitive to visible light (400 – 1000 nm). To observe the Sun in the EUV, a 150 μm thick scintillator phosphor coating has been deposited on its sensitive surface, as shown on Figure 1 (left). This coating absorbs EUV photons and re-emits visible photons, resulting in a ratio of one electron generated in the detector per EUV photon incident the coating (Halain *et al*., 2010). Figure 1 (right) shows a typical SWAP image acquired with a ten-second integration time and post-processed on-ground.

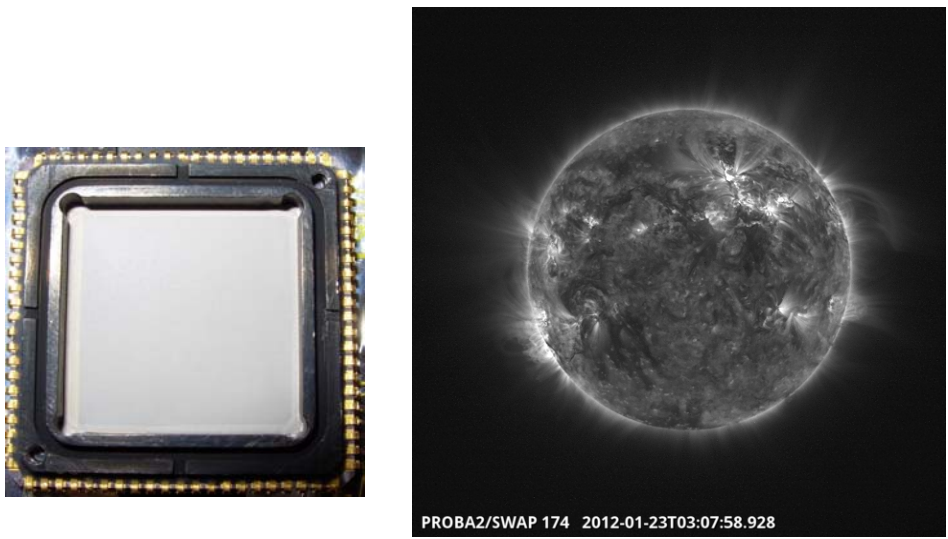

Figure 1: Left : SWAP APS detector (1k x 1k of 18 μm pixels) with scintillator coating deposited on its sensitive surface. Right: Sun image acquired by SWAP at 17.4 nm on 23 January 2012.

The use of both an APS and scintillator coating for long-term scientific observations in orbit requires careful calibration. This is necessary not only to guarantee the science return of the instrument, but also to derive some lessons for future similar instrument based on CMOS-APS sensors, such as the *Extreme EUV Imager* (EUI), Halain *et al*. 2010, onboard the ESA Solar Orbiter mission.

The SWAP instrument was intensively tested and calibrated before launch (Defise *et al*., 2007, De Groof *et al*., 2008, Halain *et al*., 2010). Major measurements and outcomes of these calibration campaigns were:

- an end-to-end instrument photometry measurement in some locations of the instrument field of view
- an end-to-end instrument spectral response,

- a tuning of the instrument parameters to optimize the image signal-to-noise ratio,
- a better understanding of the detector behavior and of image artifacts as a result of its unique capabilities and functions.

A detailed description of the SWAP instrument and the most important results from pre-flight testing appears in Part I of this article (Seaton *et al.*, 2012). Here, in Part II, we focus on the calibration and lessons learned during in-flight testing of SWAP.

We discuss the results of regular in-flight calibration campaigns to measure degradation, dark-current, straylight, and several other properties of the instrument.

## 2. Detector Performance

The SWAP detector is the key component of the instrument, and it is essential to understand its in-flight behaviour. The four major detector performances have thus been analysed, including their evolution to determine potential ageing:

- The detector dark-current
- The detector response
- The detector linearity
- The detector hot and spiky pixels

### 2.1. Detector Noise

#### 2.1.1. Types of Noise

SWAP's CMOS-APS sensor is affected by noise contributions from three primary sources: the shot noise, the spatial fixed pattern noise (FPN), and the read noise that are added in quadrature, as per Equation (1) where σ is the noise variance, to have the total noise level.

$$\sigma_{TOT} = \left[ (\sigma_{FPN})^2 + (\sigma_{SHOT})^2 + (\sigma_{READ})^2 \right]^{\frac{1}{2}} \quad (1)$$

Shot noise is a spatial, and temporal, random noise proportional to the square root of signal level. The FPN is a temporally constant non-uniform spatial noise (*i.e.* non-random) due to small differences in the pixel charge collection. The read noise is composed of two major contributors: the dark-current (DC), that results from parasitic charges that vary from pixel to pixel and are temperature dependent, and the temporal dark noise that includes all other noises that are not signal dependent.

Table 1 lists the noise contributions as measured during on-ground calibration of the SWAP flight instrument and the DC that was identified as dominant, with typical values of about 230 electrons (or approximately 7 DN, as conversion factor is 31 e$^-$DN$^{-1}$, Seaton *et al* 2012) for nominal images with a ten-second integration time.

Table 1: The major noise contributors to the SWAP images, as measured during the on-ground calibration campaign, are the fixed pattern noise (FPN) and the temporal noise, expressed in electrons, and the dark-current (DC) in electrons per second.

| Noise (Mean value – NDR mode) | Cold (-1 °C) | Hot (+20 °C) |
|---|---|---|
| FPN (spatial) | 22.97 e$^-_{rms}$ | 80 e$^-$ |
| Dark noise (temporal) | 38.3 e$^-$ | 56 e$^-$ |
| Dark-current | 19.77 e$^-$s$^{-1}$ | 230 e$^-$s$^-$ |

### 2.1.2. Dark-current

Since the rate of dark-current accumulation is a function of detector temperature, and SWAP's passive cooling system means that the detector is not kept at a constant temperature, dark rates vary widely from image to image depending on temperature changes due to PROBA2's location and orientation in its orbit and seasonal variations as a result of the Earth's orbit around the Sun.

The SWAP detector temperature is however generally kept at temperatures near 0 °C — somewhat higher than the anticipated operational temperature before launch — due to the spacecraft temperature, which is about 15° C higher than expected.

In particular, the spacecraft temperature had a very high variability over the first 18 months of the mission (the SWAP detector temperature ranging between -8° C and +10° C), but now has stabilized due to a slight change in PROBA2's orientation as it orbits the Earth.

Figure 2 shows the median and mean signal over all pixels for a randomly selected subset of all ten-second darks obtained throughout the whole mission plotted as a function of temperature. The error bar (plotted, for simplicity, as the shaded band surrounding the data points) corresponds to the distance between the location of the peak of the corresponding image histogram and the location where there are half as many pixels in each bin as there are at the peak itself.

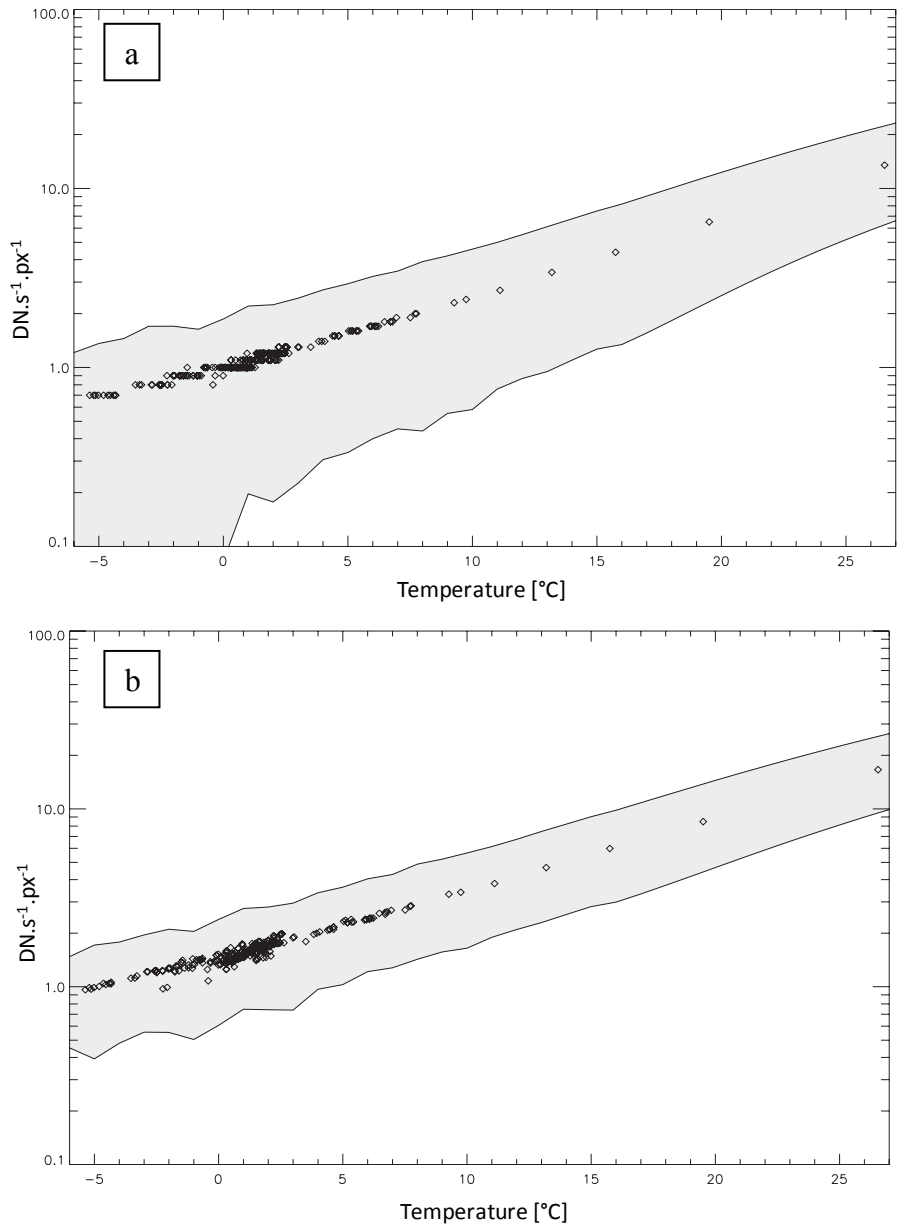

Figure 2. SWAP median (a) and mean (b) signal in all pixels for a randomly selected subset of all ten-second darks obtained throughout the whole mission plot *vs.* temperature, the shading representing the error bar plotted relative to the data points.

The non-uniformity of CMOS pixels' dark response is responsible for the large difference in mean and median image noise, and for the fact that the error bar in the median plot is so wide that it cannot be displayed properly in the plot (and thus runs off the edge of figure). Since each pixel in a CMOS detector has its own electronics, each pixel indeed behaves as an independent detector. As a result, the dark-current in each pixel of a single image varies widely. At SWAP's nominal operational temperatures, near 0 °C, the dark rate for the vast majority of pixels is below 1 $DN.s^{-1}$. However, dark-current accumulates much more rapidly in a small number of outliers, leading to a large increase in overall image noise even in the domain where most pixels are relatively noiseless.

Using the conversion factor 1 DN= 31 e$^-$ (Table 2 in Seaton *et al.*, 2012), the 1 DN.s$^{-1}$ corresponds to 31 e$^-$.s$^{-1}$ dark-current measured in flight, which is roughly 50% higher than the 19.77 e$^-$.s$^{-1}$ measured at -1 °C before launch (see Table 1). The reason for this increase after launch is most probably due to initial aging of the detector during the integration and first months in space and/or a limited number of pixels dominating the statistics as explained before.

### 2.1.3. Noise Evolution

Calibration sequences are performed on a regular basis of one or two weeks since the instrument switch-on in November 2009. These sequences include a series of five images with three-second integration time captured in dark conditions.

The SWAP instrument is operated without a shutter. To avoid direct illumination by the Sun during calibration-image acquisition, the in-flight calibration sequences are performed while the spacecraft is pointed away from the Sun by a few degrees (3° is sufficient to ensure that no direct EUV light reaches the detector and that any additional internal reflections are sufficiently attenuated, as shown in by the in-flight straylight analysis presented later in this article).

We monitor the overall evolution of detector dark-current by computing the average value over the entire sensor area of the five dark raw calibration images (that is, images with no correction or post-processing). To further suppress noise, we average the results of the individual images into a single value.

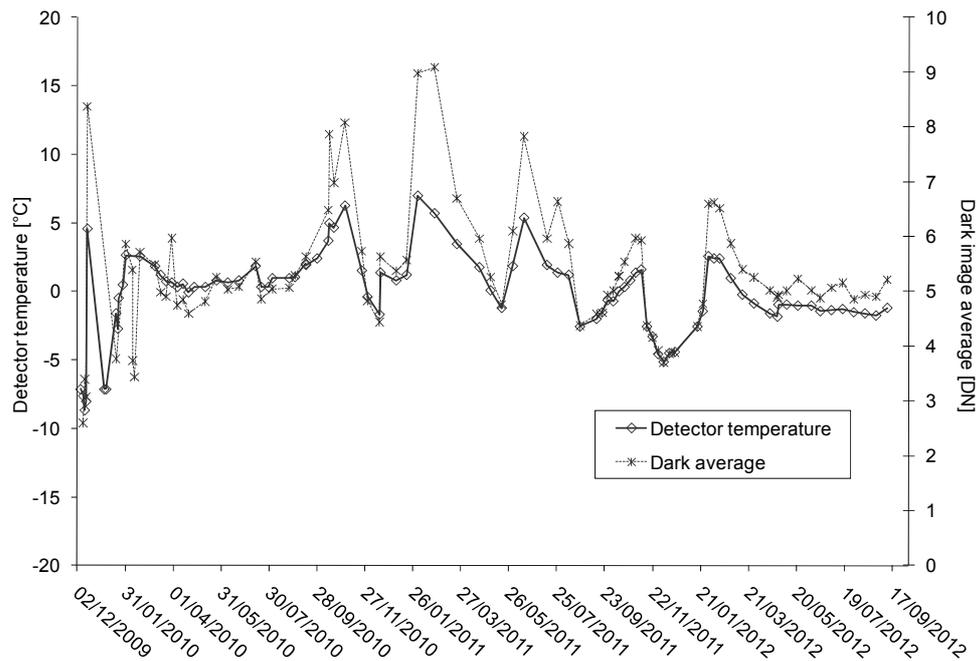

Figure 3: SWAP detector temperature evolution since launch (left axis) compared with dark image average (right axis, computed from raw images and expressed in digital number for three-seconds exposure time).

Figure 3 shows the temporal evolution of the average signal in dark images obtained by SWAP, measured at roughly regular intervals since launch, together with the SWAP detector temperature, which closely tracks the spacecraft temperature variation including seasonal effects. It is clear that the detector dark average closely follows the temperature evolution, indicating the temperature-dependent dark-current dominates the dark images.

**2.1.4. Dark-current Correction**

In typical astronomical imaging, dark-current can be removed from images straightforwardly, by subtracting dark images – images constructed by averaging a series of observations made with a closed shutter that therefore contain only noise – from the data images. Since this noise varies with temperature, it is essential that these dark images are obtained with the detector at the same temperature as the corresponding observations. Since most instruments are actively cooled and kept at constant temperature, this is rarely a problem for anyone attempting to calibrate their observations.

However, for SWAP this procedure is complicated in two ways:

- First, because SWAP does not have a mechanical shutter, obtaining measurements of dark-current requires us to slew the spacecraft away from the Sun to a region of the sky where EUV emission and straylight is minimum.
- Second, because SWAP's temperature is variable over the course of even a single orbit, dark-current varies from image to image as well. Although it would be possible, in principle, to obtain dark frames whenever a significant temperature change occurs on the spacecraft, limitations on both telemetry and spacecraft maneuvers make this impractical.

Additionally, SWAP calibration software is distributed via the IDL software package *SolarSoft* so users of SWAP data can prepare their own files on local computers (see Freeland and Handy, 1998, for a discussion of *SolarSoft*, and Seaton *et al.*, 2012, for a discussion of the SWAP calibration software). As a result, it is also impractical to maintain and distribute a large library of pre-processed dark images to cover a range of temperatures.

Instead, dark images are generated on the fly during SWAP calibration using an empirical model based on dark-current observations obtained during a series of calibration campaigns that have been conducted throughout the PROBA2 mission.

Thus it is possible to remove dark-current from SWAP images obtained at any temperature that has been recorded during the PROBA2 mission.

In order to generate this empirical model, we used approximately 3000 dark images obtained during calibration campaigns run over the course of a year. These campaigns typically are run at nominal detector temperatures, but are supplemented by darks obtained during the cooling-down phase that followed a detector bake out. The addition of these relatively high-temperature darks gives us a data set that reveals dark-current variation across the detector for temperatures ranging from about –2 °C to +36 °C.

To reduce the effects of both shot noise and sampling error in the individual dark images, we grouped images into 0.25 °C temperature bins and computed the mean dark noise as a function of temperature for each pixel. At high temperatures, where we had relatively few images, many bins did not contain data. However, since the effects of both shot noise and sampling error decrease for larger signal levels, these few high-temperature values are sufficient for the construction of our model.

We then fit the data pixel-by-pixel with a third-degree polynomial and store the coefficients for each pixel. Figure 4 shows the binned data and fit for a representative pixel. In general this method allows us to reconstruct dark-current for each pixel as a function of temperature to within just a few percent for the large range of possible detector temperatures that occur during the course of a year. Figure 5 shows an example of how images are corrected with this method.

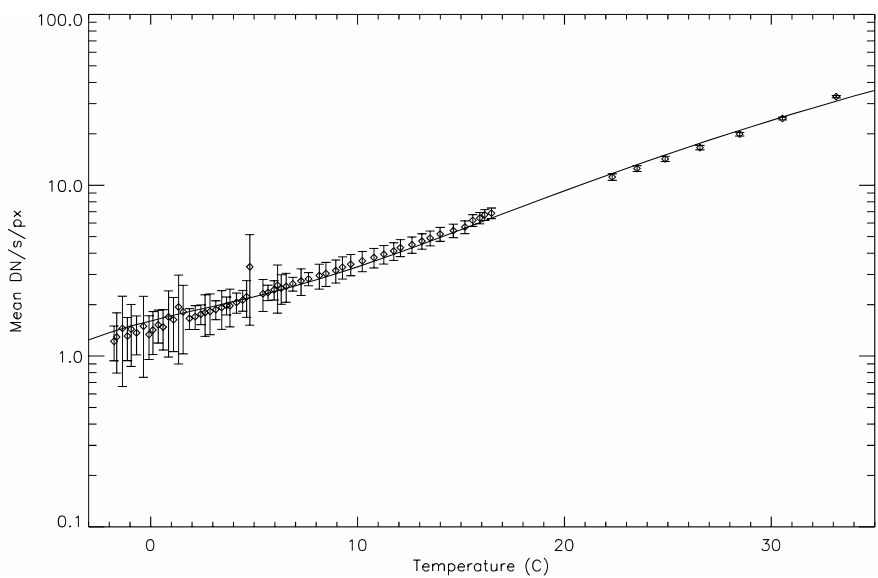

Figure 4: Measured and modeled dark-current in a representative SWAP pixel. The error bars are largest for low temperatures largely because quantization error is most significant for the very small corresponding values.

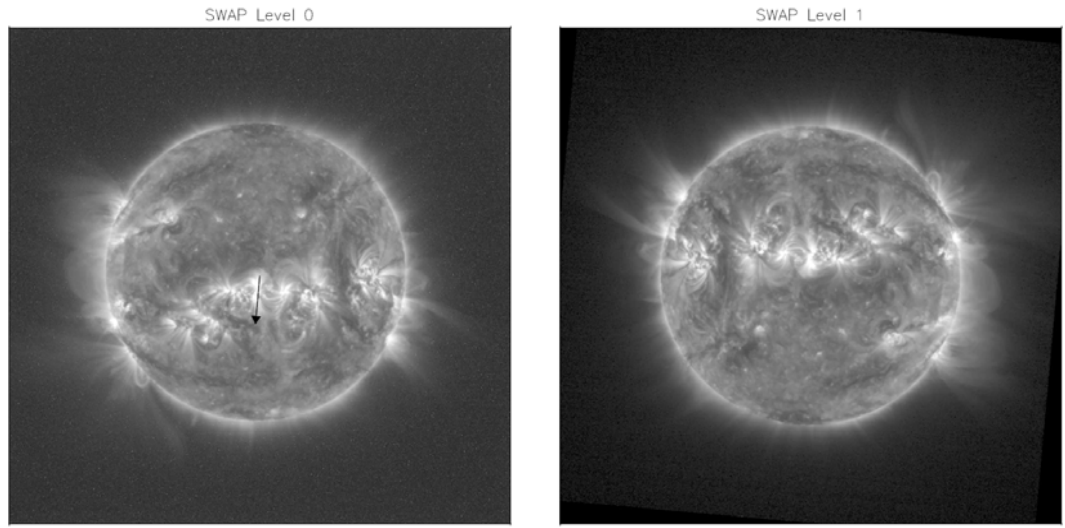

Figure 5: An original SWAP image as observed onboard (*e.g.* Level-0, left) and the same image after correction for spacecraft orientation, pointing, and instrumental effects such as dark-current (*e.g.* Level-1, right). The arrow in the uncorrected image indicated the direction of solar North (North is oriented directly upwards in the Level-1 image).

## 2.2. Detector response

The detector response is monitored with regular images of a near-UV LED. As it is however difficult to make a distinction between LED and detector degradation, a comparison with a calibrated external instrument is used to demonstrate that the detector degradation is not the major contributor to the observed detector-response evolution.

### 2.2.1. Response monitoring

We monitor the in-flight detector response of SWAP using two current-driven near-UV LEDs that emit at 500 nm (which we refer to as LED-A and LED-B). Since the detector coating is transparent to the near-UV, the LED wavelength was selected to correspond to the range of sensitivity of the detector without the need for the scintillator coating.

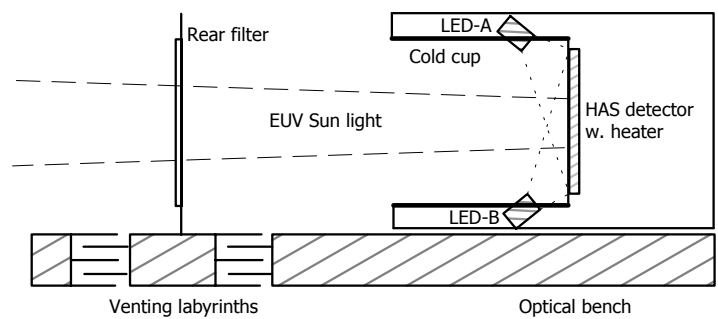

Figure 6: SWAP focal plane assembly and detector cavity (distance between rear filter and detector is ≈ 8 cm).

The two LEDs are located in the vicinity of the detector, behind the rear filter assembly, as shown in Figure 6 (Halain *et al.*, 2010). In Figure 7, which shows the both LED-A and LED-B pattern on the detector, it is clear that the LEDs unfortunately do not provide uniform illumination across the detector and, furthermore, that LED-B is partially vignetted. Nevertheless, the combination of the two patterns is sufficient to characterize the detector response, as all pixels are adequately illuminated either by one or the other LED. However, because the illumination is non-uniform, and because the LEDs themselves are only at the end of the optical path, LED images cannot be used directly for flat-field correction and gain calibration of the detector.

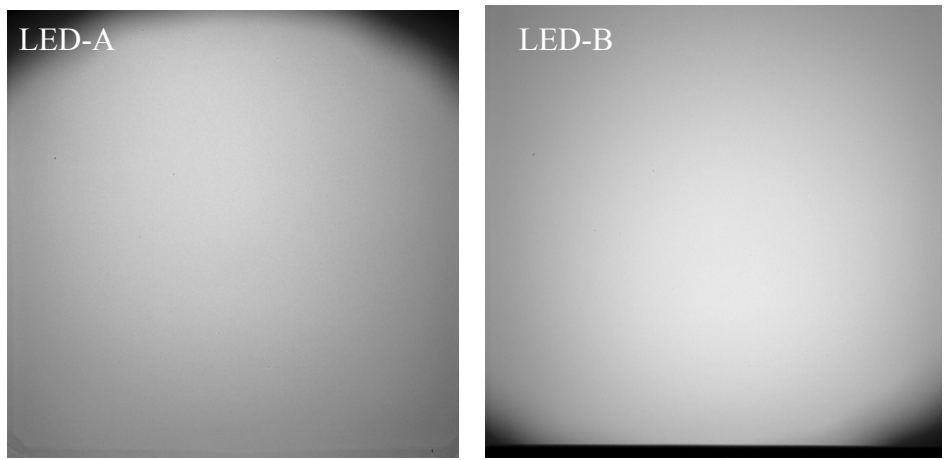

Figure 7. SWAP images of the two UV calibration LED-A (left) and LED-B (right) that are located in the vicinity of the detector (approximately at 3 cm from the detector center).

Series of five LED-A and five LED-B images with 3 second integration time are captured in dark conditions over the same calibration period as for dark calibration images. The response of LED is obtained by the same method as for dark-current evolution analysis.

Figure 8 shows image averages for both LEDS, plotted together with the dark image average. This figure suggests that the detector-response has declined by 4% over the first two years of the PROBA2 mission. Linear fitting of the results shows a clear decoupling of the averaged response to LED illumination and of the averaged dark images, indicating that the detector response evolution is not likely linked to the detector temperature. As a consequence, this decline must be the result of overall degradation of either the detector sensitivity and/or the LEDs. However, since there is no way to measure the absolute brightness of the LEDs on SWAP, it is impossible to determine from these measurements whether the detector or the LEDs themselves are responsible for this overall decline.

On shorter timescales, as shown in Figure 9, where we have plotted the LED-A image average together with the detector temperature, it is clear that the detector response to LED illumination is correlated with the detector temperature. Since the LEDs are in close proximity to the detector, it is very likely that the LED temperature follows the detector temperature very closely. So it is clear that the short-term variation in observed brightness is linked to emissivity variation of the LEDs as a function of their temperature. Since it is unlikely that LEDs or detector become better performing with ageing, the result of Figure 8 shows at least that the SWAP degradation is certainly less than 2% per year.

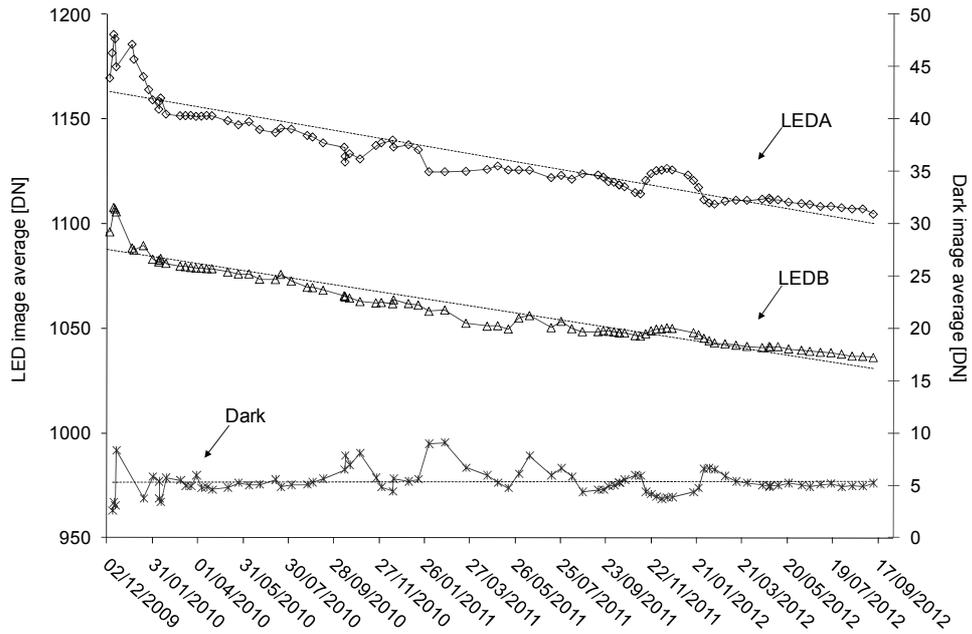

Figure 8: SWAP calibration LED image average (left axis, computed from raw images and expressed in digital number) compared with dark image average (right axis, computed from raw images and expressed in digital number for three-second exposure time).

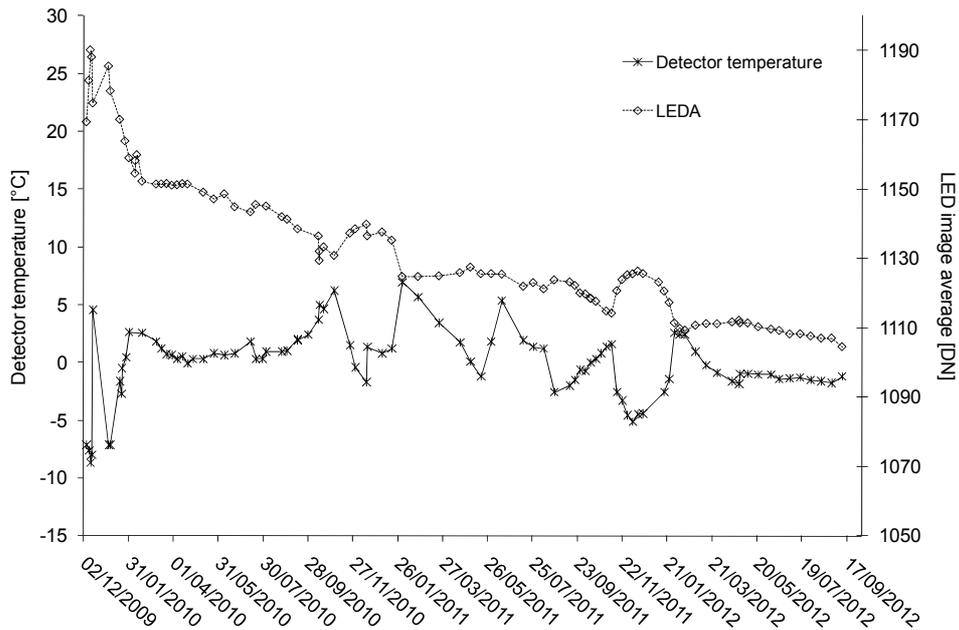

Figure 9: SWAP detector temperature variation since launch (left vertical axis) compared with LEDA image average (right vertical axis, computed from raw images and expressed in digital number for three-second exposure time).

### 2.2.2. Response Comparison

In order to help separate LED evolution from detector evolution, it is necessary to compare SWAP observations to those of a well-calibrated, external source.

To do this, we compared the evolution of integrated SWAP response over the course of the mission to calibrated Level-2 (version 2) spectra from the *Extreme-Ultraviolet Variability Experiment* (EVE) on the *Solar Dynamics Observatory* spacecraft.

We computed the average of EVE spectra obtained over the course of an hour, between 0:00 and 1:00 UT, and converted these average spectra from units of energy flux to photon flux (Figure 10) – that is, from $W.m^{-2}$ to $ph.s^{-1}.m^{-2}$ – since SWAP cannot distinguish between photons of different wavelengths within its passband. We then modulated these spectra using the SWAP wavelength-response function (Seaton *et al.*, 2012) and integrated the resulting spectra to obtain a synthetic SWAP integrated intensity per pixel (Figure 11).

We compared this value to the average of the measured SWAP integrated intensity, which we refer to as *SWAVINT* (corresponding to its keyword in SWAP FITS files) for the corresponding hour, corrected for seasonal variation in earth-sun distance. The evolution of both synthetic and real SWAP response (*SWAVINT*) is shown in Figure 12. We found that, up to a time-invariant factor of about 1.11, the synthetic and real values are almost perfectly coincident throughout much of the 680-day period that we studied.

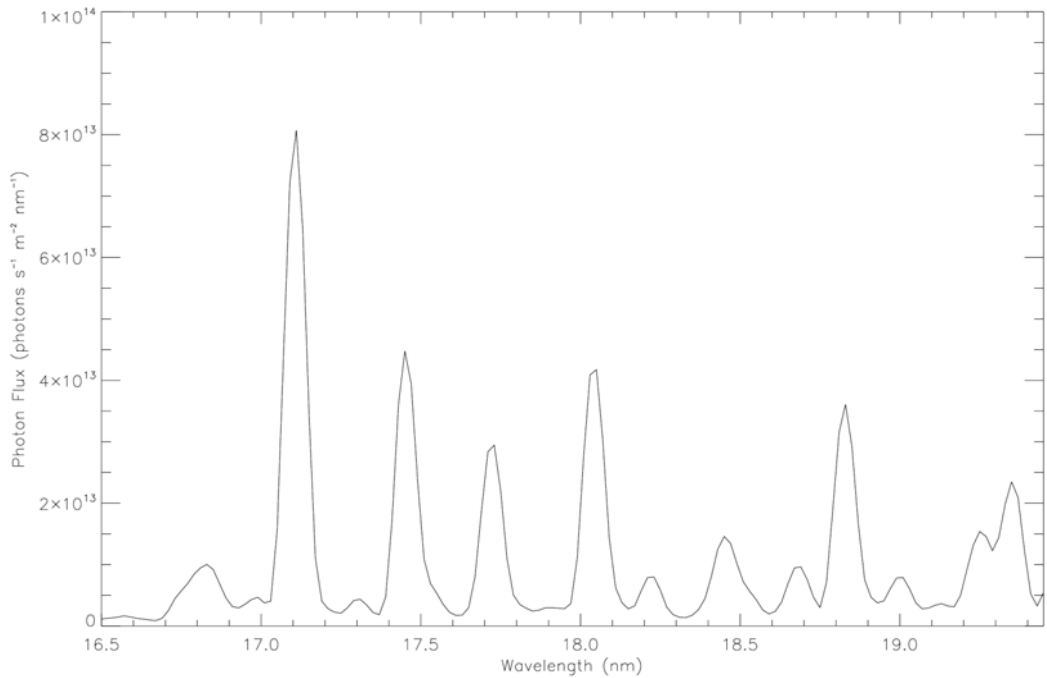

Figure 10: EVE irradiance converted from units of energy flux ($W.m^{-2}$) to photon flux ($ph.s^{-1}.m^{-2}$).

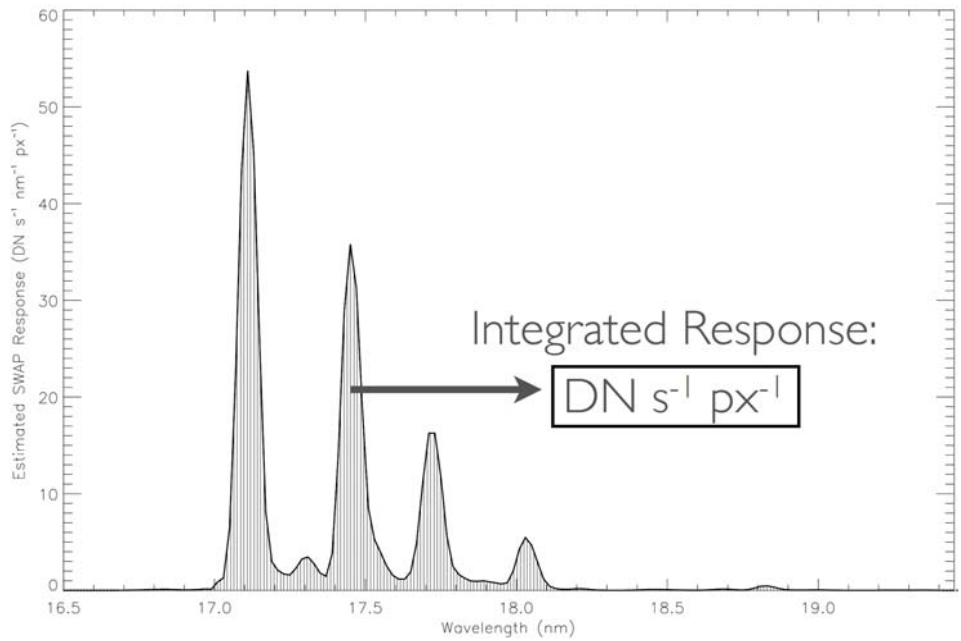

Figure 11: The estimated SWAP response is obtained by modulation of the EVE photon flux by the SWAP bandpass, and then integrated to have a synthetic SWAP integrated intensity per pixel.

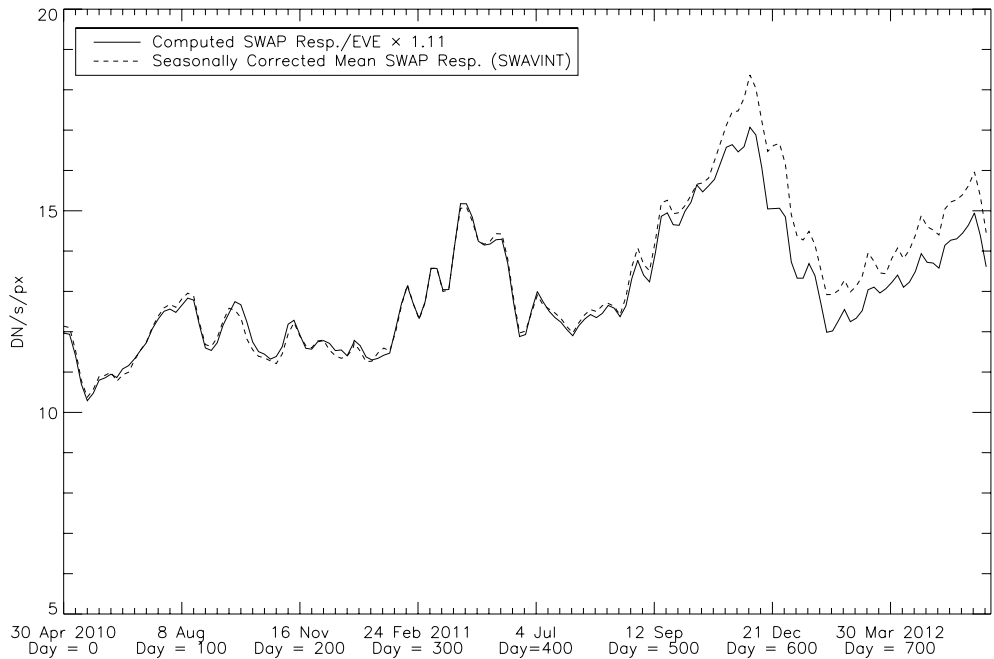

Figure 12: Temporal evolution of SWAP integrated intensity (blue) compared to synthetic response generated from EVE spectra (black) showing good correlation for most of the SWAP mission.

Thus we can conclude that SWAP has degraded no more than EVE over this period. Further, since it is unlikely that EVE, a spectrometer, and SWAP, an imager, would degrade at exactly the same rate, this correlation suggests SWAP has not measurably degraded during the first two years of its mission.

The roughly 10% mismatch in predicted and real results is likely the result of the combination of two factors. First, it is possible that the laboratory-measured

SWAP response function contains some error and is undervalued slightly. Second, EVE's periodic calibration rocket flights demonstrate that EVE is degrading in an expected and correctable manner but there is some uncorrected degradation in the EVE spectra which thus have some uncertainties (Hock *et al*., 2012), which could result in an underestimate of photon flux by a few percent.

As highlighted in Figure 13, the synthetic signal begins to diverge from SWAVINT around November 2011. One possibility is that this divergence is the result of an uncorrected degradation in EVE.

It may also be evidence of a more fundamental change in the coronal brightness to which both instruments respond differently. For example, since the corona near 17.4 nm is brightening along with the rise phase of the solar cycle, there may be new areas of coronal brightness appearing outside of one of the two instruments' fields-of-view or an increase in instrumental saturation due to increased occurrence of bright coronal structures. Either of these phenomena would have same effect as changing the input spectrum observed by one instrument or the other and could cause a divergence of the two instrumental response functions.

Nonetheless, the general conclusion is that SWAP has not significantly degraded during roughly the first two years of its mission, and the decline in LED calibration brightness is more than likely due to changes in the LED lamps themselves.

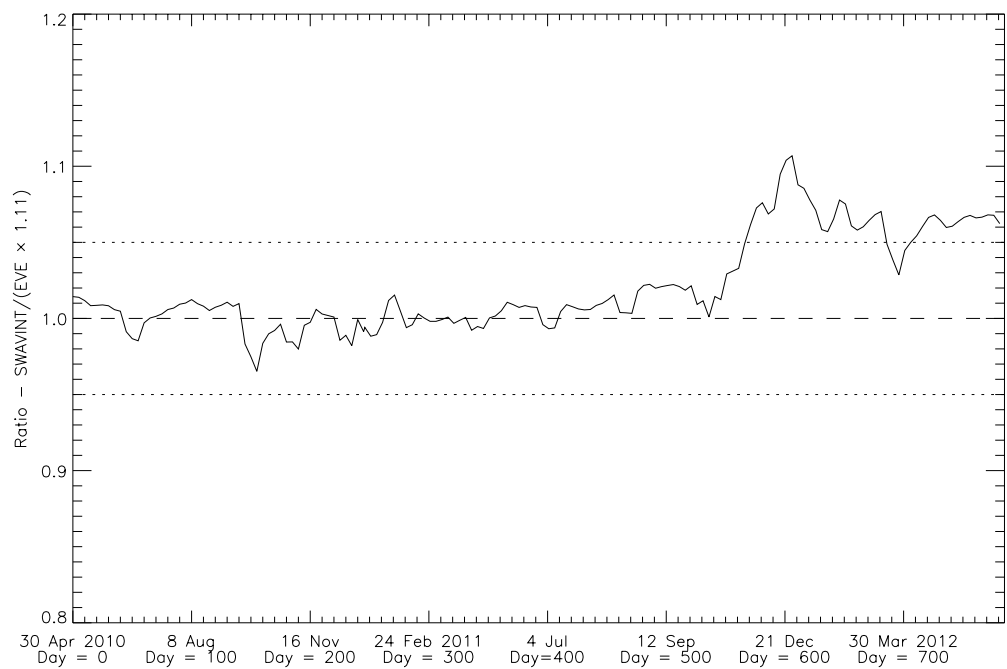

Figure 13: Temporal evolution of the ratio between SWAP and EVE integrated intensity, which shows good correlation until approximately November 2011.

Another indication that the SWAP sensor, or its coating, has not degraded in a significant way is to look at those locations in the images where the EUV exposure has been the strongest, *i.e.* at the solar limb. This is shown in Figure 14, where a slight degradation can be noticed but the resulting degradation accounts for only a very small reduction of signal, on the order of 0.1%. We thus conclude again that SWAP has not degraded in any significant way. Since this amount of degradation is far below the noise levels in most SWAP images, it is undetectable in nominal observations and is not significant enough to be seen in the SWAP/EVE comparison.

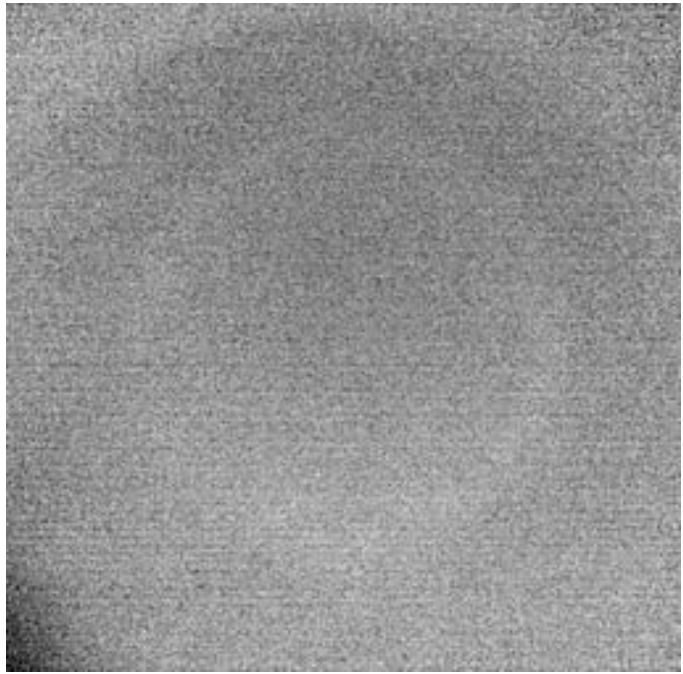

Figure 14: Ratio of three-second LED images from 2012 to similar images obtained around the beginning of the PROBA2 mission in 2010, showing a the results of a slight 'burn-in' ring in the region of maximum coronal brightness near the limb of the Sun (image is here the full detector size of 1k x 1k pixels).

## 2.3. Linearity

We characterize the detector linearity by comparing the signal density [in DN] of all the pixels for two images of different exposure duration.

In the corresponding density plot obtained from three seconds *vs.* 5 seconds LED images, shown in Figure 15, each point is defined by the pair of DN values a pixel gathered in the weakly exposed image (*x*-coordinate) and strongly exposed image (*y*-coordinate). The density of each point in the map is given by the color scale: red points correspond to a large density of pixels with the same DN pair, and blue indicates a low pixel density. In a perfectly linear detector, this density would result in an entirely linear feature in the plot. However, because the detector is nonlinear, the best fit deviates from the idealized linearity curve (a line) as the signal level increases. Only non-saturated images are used for this density plot, to avoid a spread of the pixel values on the top of the graph due to pixels that are not saturated in one image but that reach saturation in the other, which then fall far below the expected curve.

Comparing images obtained in December 2009 to images from July 2010 reveals a 5% non-linearity with a 0.5% evolution over six months.

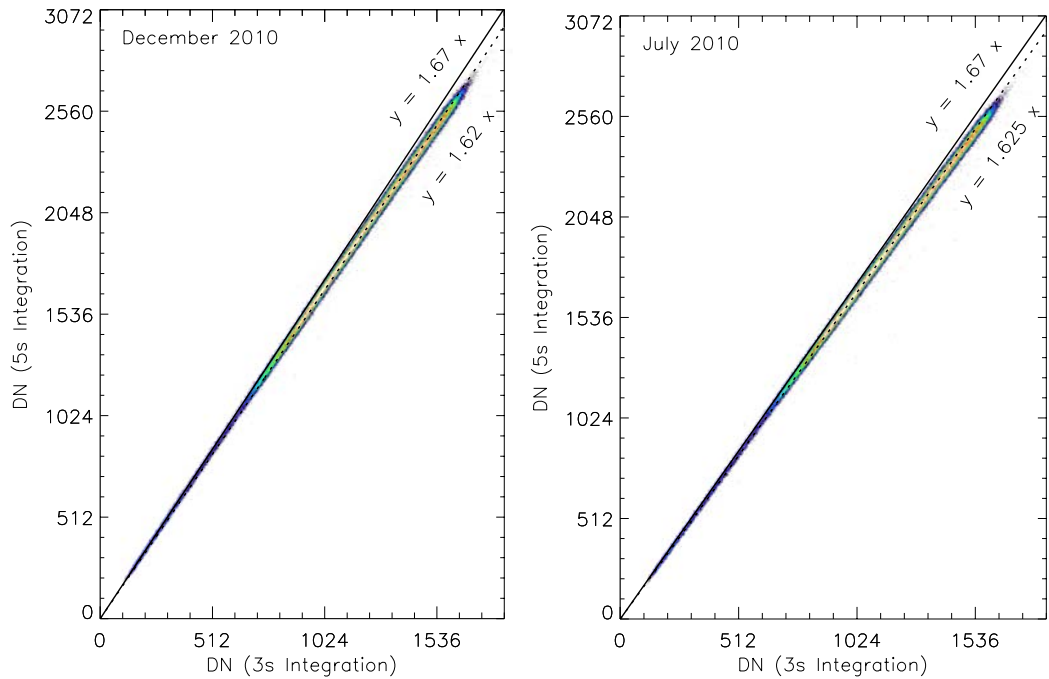

Figure 15: Density plot of pixel brightness for three-second *vs.* five-second images of the LED-A, obtained in December 2009 (Left), best fit *y* = 1.62 *x*, and in July 2010 (right), best fit *y* = 1.625 *x*. The *y* =1.67 *x* slope corresponding to linearity is drawn for comparison. The scale is limited to 3072 for clarity.

The similar density plot of Figure 16 gives the brightness of Sun in all the pixels for two images, one with a five-second integration time versus one with a ten-second integration time, captured consecutively to control for evolution of the solar signal itself. The best linear fit of the pair is given by *y* = 1.95 *x*, while the nominal linearity curve is *y* = 2 *x*, also indicating a 5% non-linearity of the detector.

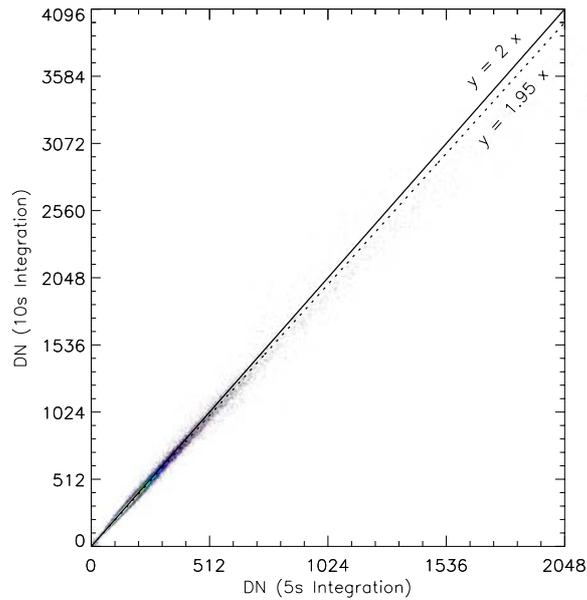

Figure 16: Density plot of pixel brightness for five-second *vs.* ten-second images of the Sun, captured in July 2010.

## 2.4. Hot and Spiky Pixels

A "hot" pixel is a pixel having an unusually high value, even in dark condition, related to a defect in the detector array. As shown on Figure 17, the number of pixels in raw dark images whose value is higher than 2048 DN, 1024 DN, and 512 DN is very low as compared to the total number of pixels (1k x 1k), and has not increased significantly since launch.

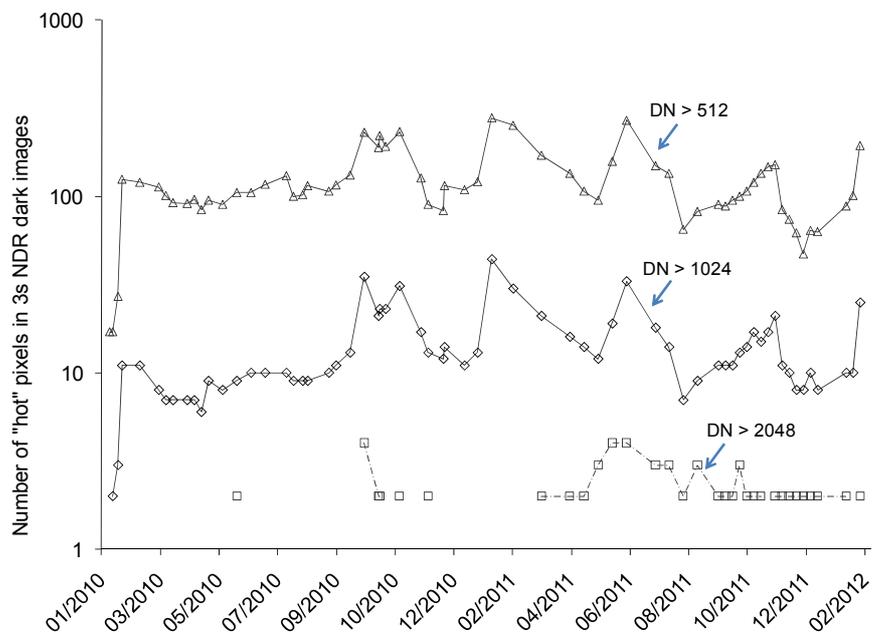

Figure 17: Number of pixels above a certain threshold (512, 1024, and 2048 DN) computed from dark image sequences captured since launch.

As Figure 18 shows, there also seems to be a correlation between the number of hot pixels in dark images and the detector temperature. This indicates that many of the so-called hot pixels are hot because of an excess dark-current. Since dark-current increases as a function of temperature, the number of hot pixels also increases when temperature increases.

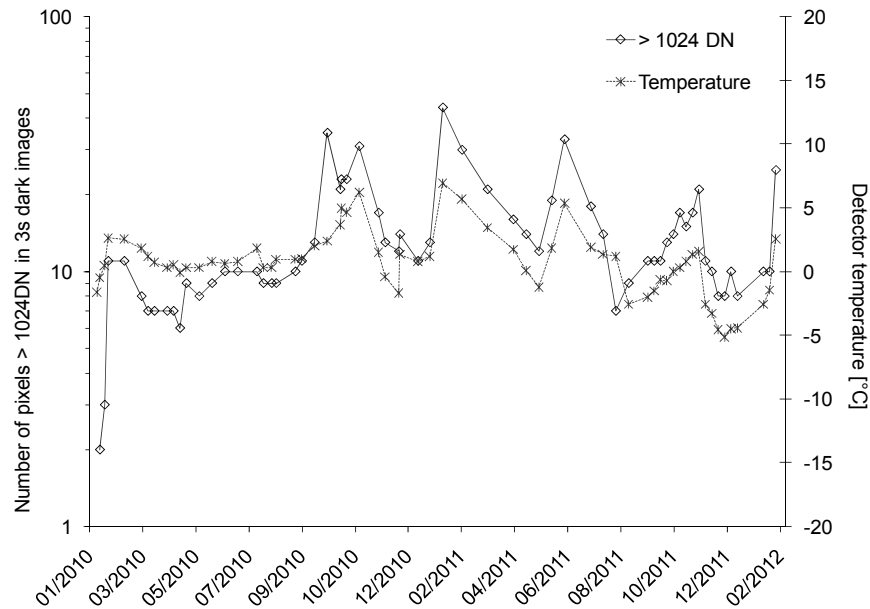

Figure 18: Number of pixels above 1024 DN threshold computed from dark-image sequences as compared with detector temperature.

However, the rate of hot-pixel detections in nominal SWAP data images (Figure 19) tends to increase at a rate of ≈ 6100 additional hot pixels per year, showing that the detector is degrading at significantly less than 0.5% per year.

The difference of hot-pixel evolution between dark and nominal images can be explained by the difference in the approaches used to identify hot pixels in dark and nominal images.

In the nominal images, we search for pixels that are much brighter (or darker) than their local neighborhood, indicating that there is a spike. In dark images we search for pixels over a certain threshold. The nominal-image routine is consequently more sensitive to pixels that are just a little bit too bright. It can however not be applied on dark images because the darks are highly non-uniform compared to nominal images, where the solar signal dominates detector noise in most places. When there is only noise, the nominal-image routine would thus identify many normal pixels as spikes just because they happen to be a small amount brighter than their local neighborhood.

The rate of detections of hot pixels is however related to the quality of dark-current correction. The better dark-current is corrected, the fewer hot pixels are likely to be detected since each pixel's behavior is more or less completely independent from its neighbors. In December 2010 an improved dark-current model has been implemented that significantly reduced the number of detected hot pixels as can be observed on Figure 19.

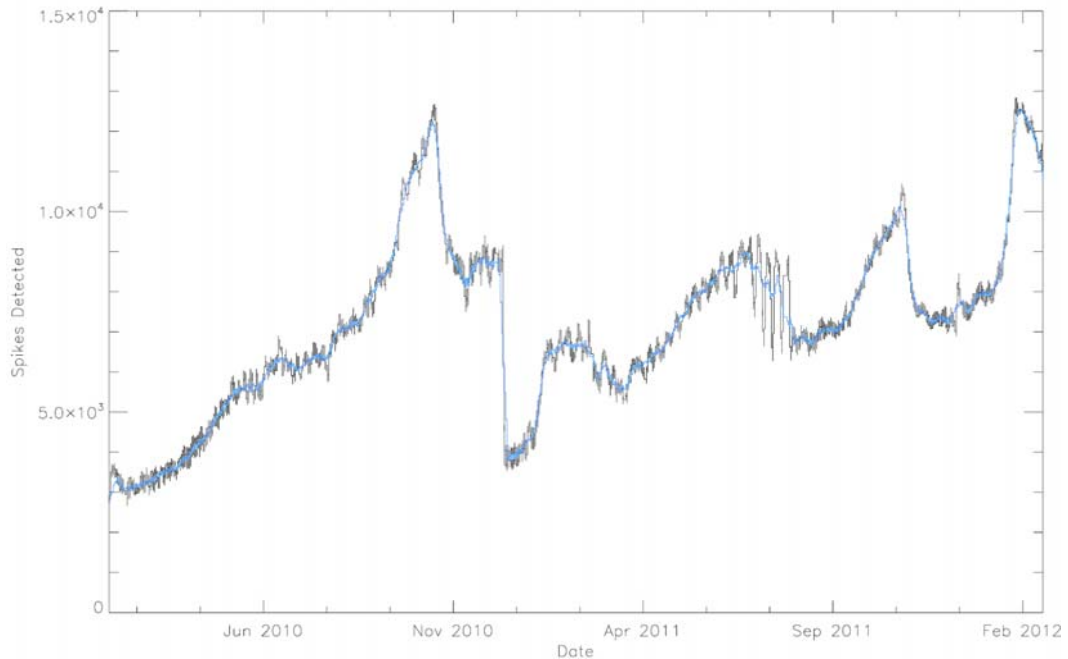

Figure 19: Rate of hot-pixel detection in nominal SWAP data images. The drop of hot-pixel number at beginning of 2011 is related to an improved dark-current-removal image processing.

In addition to the hot pixels, some images contain "spiky" pixels, which correspond to pixels that are activated by interaction with the trapped particles in the magnetosphere. The NASA AP-8 and AE-8 models of trapped protons and electrons given by the Space Environment Information System (SPENVIS: Heynderickx, 2002) are shown in the maps in Figure 20 and Figure 21.

The software that generates Level-1 science images identifies these pixels and replaces them with the median value of their neighbors. By counting how many bright pixels this algorithm identifies and mapping them with respect to the satellite location at the moment of the image acquisition we can identify regions of the magnetosphere where the level of trapped energetic particles is unusually high. To make such maps we identified the images where the number of spikes detected is six or more standard deviations above the series of spike counts from images acquired around the same time. The density of these images' locations is mapped in Figure 22. Comparing this figure to the SPENVIS maps, we see that this procedure has recovered the location of the South Atlantic Anomaly (SAA) and auroral ovals. Comparing the SWAP map to the older SPENVIS maps also reveals the known westward drift of the SAA (Badhwar, 1997).

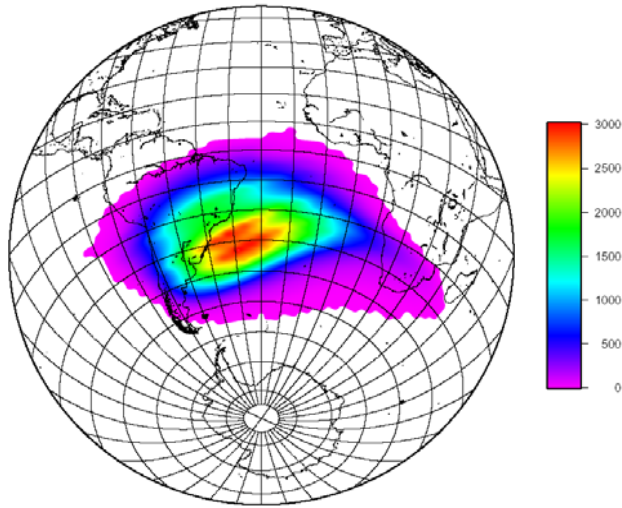

Figure 20: AP-8 MAX model of the flux of trapped protons with energies above 50 MeV during solar maximum at 725 km altitude (number of particles.cm$^{-2}$.s$^{-1}$).

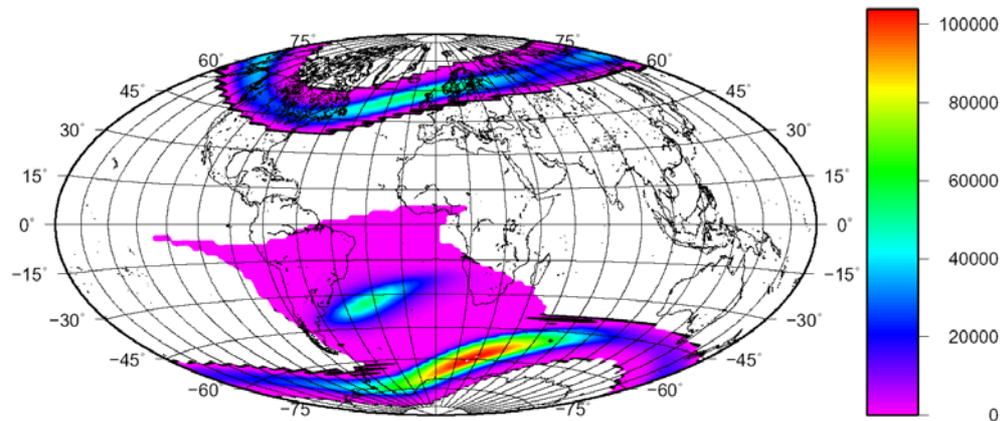

Figure 21: AE-8 MAX model of the flux of trapped electrons with energies above 1 MeV during solar maximum at 725 km altitude (number of particles.cm$^{-2}$.s$^{-1}$).

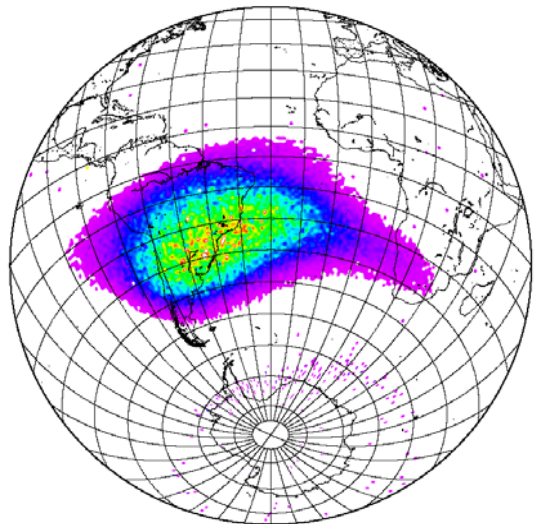

Figure 22: Locations of the PROBA2 satellite corresponding to images where the number of spikes detected is 6 or more standard deviations above the series of spike counts from images acquired close in time.

# 3. Instrument Straylight

## 3.1. Out-of-Field Straylight

We have performed measurements to characterize straylight in SWAP images both during Sun-pointed observations and off-pointed observation when the Sun is far from the field of view. Out-of-field straylight measurements ensure that the dark and LED images captured during calibration sequences (at 3º off-pointing) are not contaminated by any residual signal from the Sun.

In order to make these measurements, we obtained a series of images with integration times of both 10 and 40 seconds for a range of pointing between 0 and 180 arcmin from Sun-center. Figure 23 shows the ratio of mean values of each raw image (Level-0) average to the mean value of a Sun-centered image (that is, an image with a 0 arcmin off-point) *vs.* the off-pointing angle itself. For comparison, we plot these measured values together with a corresponding ray-tracing model curve (obtained for a light source at infinity having the same divergence than the Sun).

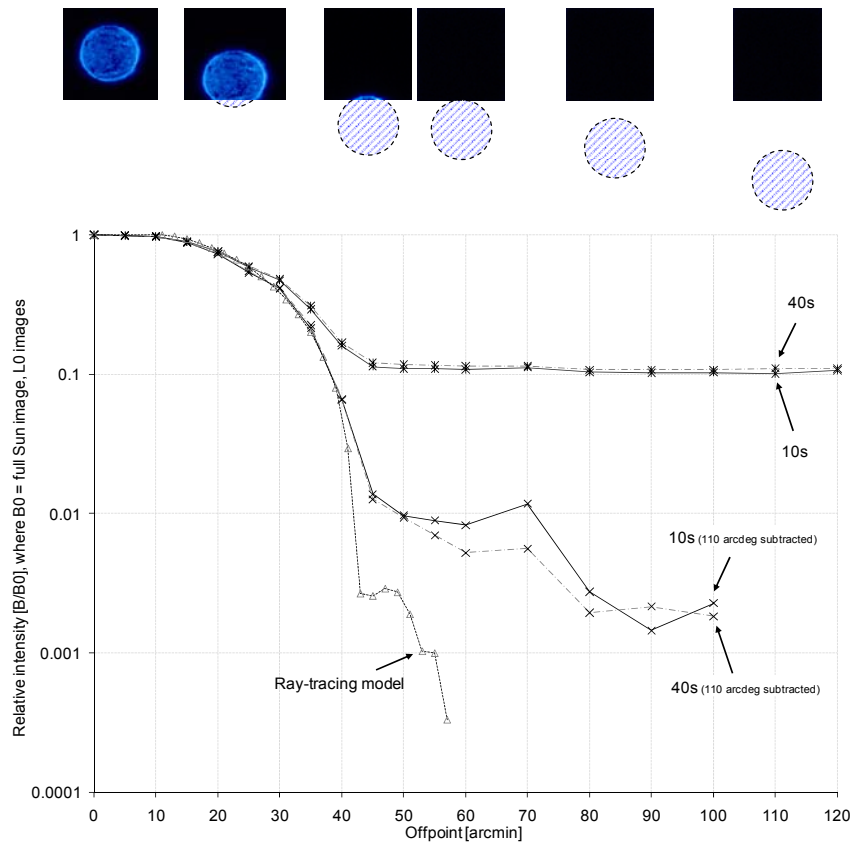

Figure 23: Normalized averages from images with integration times of 10 seconds and 40 seconds, with and without subtracting a 110 arcmin off-pointed image. The values plotted range from nominal pointing (Sun center) up to 120 arcmin off-point and are shown together with the ray-tracing model of the Sun entering the instrument. For reference, the cartoon above shows the corresponding Sun location with respect to the image frame.

In the case of the Level-0 images average, a 10% background is visible when the Sun is out of the field of view. In order to determine how much of this background is the result of straylight, we computed these averages after subtracting one of the most off-pointed images (110 arcmin) in the set. As we see in the Figure 23, there is good agreement with ray-traced values up to 45 arcmin, beyond which we believe most of the remaining signal is due to detector noise, and in particular the dark-current. It is clear that the straylight level due to the out-of-the field Sun is less than 1% of the Sun average[7].

To confirm this result, we performed the same analysis using calibrated science images (Level-1 images), in which bad pixels have been replaced by the average of their neighbors, a dark-current map has been subtracted, the effects of spacecraft pointing, orientation, and other optical effects have been corrected, and which have been normalized with respect to the exposure time (for additional information on image calibration see Seaton *et al.*, 2012).

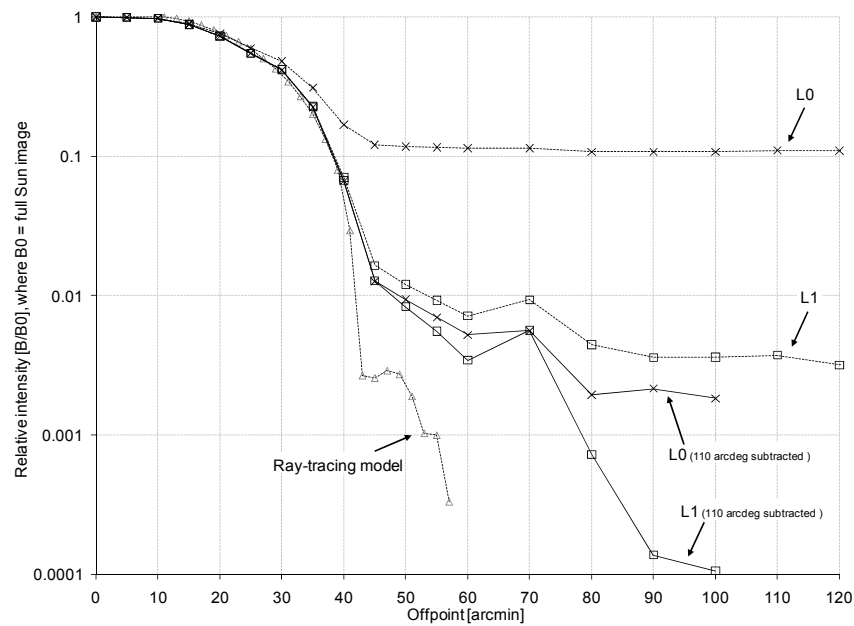

Figure 24: Normalized averages of level-0 and level-1 images with 40-second integration time, from nominal pointing (Sun center) up to a 100-arcmin off-point, with and without subtracting the 110 arcmin off-pointed image, together with the ray-tracing model of the Sun entering the instrument.

Figure 24 shows a comparison between level-0 and calibrated level-1 images with a 40 second integration time. The use of Level-1 images removes the detector dark-current from the image background and provides a better match to the theoretical curve. The effect of removing one of the largest off-pointed image (110º), before averaging the Level-1 image, then reveals the straylight contribution at the largest off-points.

---

[7] Due to in-flight image-sequence constraints, the images at 70 arcmin were captured after having pointed back to the Sun for a short duration, and are therefore perturbed by a remanent image of the Sun most probably due to the detector scintillator coating behavior (De Groof *et al.*, 2008).

## 3.2. In-Field Straylight

Separating in-field straylight from the coronal background is much more complicated than estimating the out-of-field straylight. However, during a lunar occultation that occurred as a result of the 15 January 2011 eclipse, we were able to highlight a possible straylight contribution to the nominal Sun images. By comparing the coronal brightness to the residual brightness in images in which some of the corona is obscured by the moon, we can obtain an estimate of the residual in-field straylight. Figure 25 and Figure 26 show the image values along a radial slice that passes from the Sun center to the image border where the Moon hides the Sun (difference between Sun image with and without the Moon).

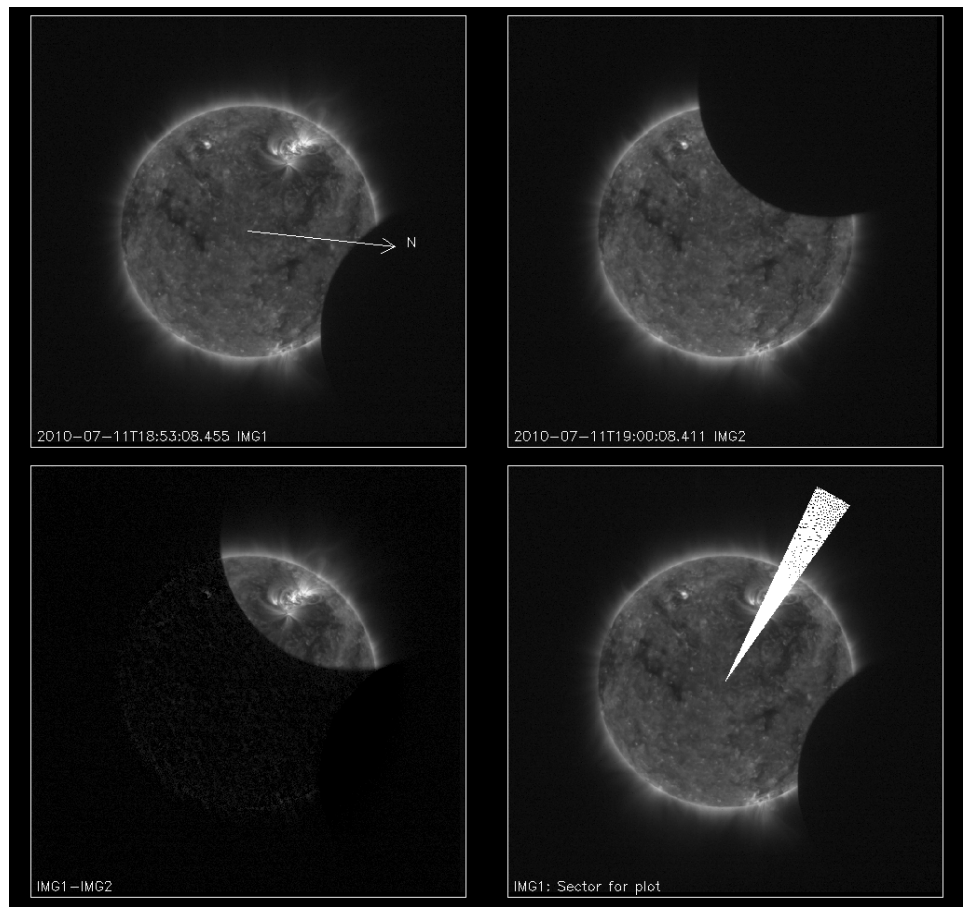

Figure 25: Straylight analysis using images of the 11 July 2010 eclipse. Top Left: An image of the eclipse indicating the direction of solar North. Top Right: A second image of the eclipse in which the moon has moved through the frame, obscuring the active region near the top of the image on the left. Bottom Left: The difference between the two images, leaving only the active region with straylight removed. Bottom Right: Image showing region where signal falloff has been sampled.

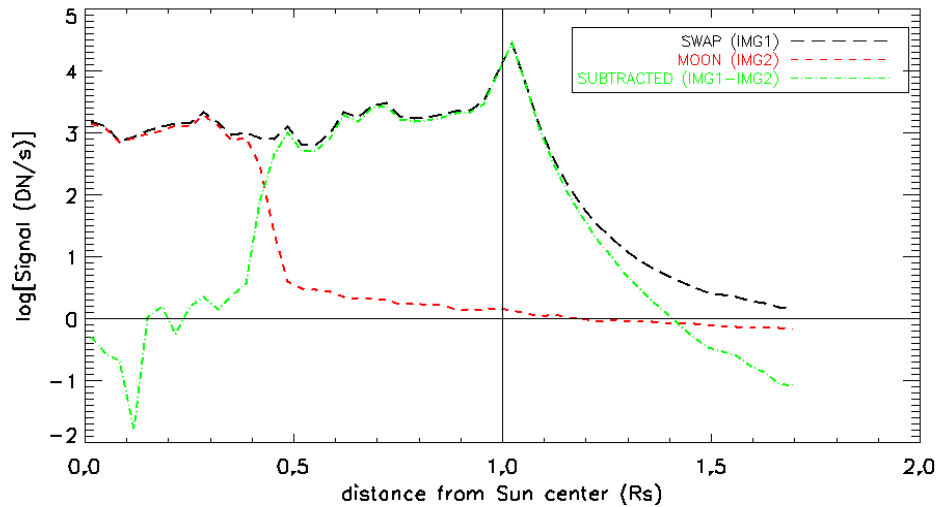

Figure 26: Brightness falloff measured during eclipse of 11 July 2008. The black curve shows the original image, while the red-curve an image where the Moon obscures part of the Sun, leaving only the instrumental straylight. The green curve shows the difference, revealing that straylight is relatively significant at heights above 1.3 solar-radii. The differences below 0.5 solar-radii are due to the solar variability (the two images were taken with a few minutes interval) and noise variation.

These figures suggest that straylight is a significant source of brightness contamination in SWAP images, especially at large radial distance where the inherent brightness of the corona is only a few $DN.s^{-1}$. At least some of this straylight is the result of the wings of the instrumental point-spread-function rather than specular reflection, and can, in principle, be corrected using deconvolution techniques (*e.g.* DeForest *et al.*, 2009; Shearer *et al.*, 2012). An effort to extend such techniques to SWAP images is ongoing, but has not been completed at this time.

### 3.3. Possible Straylight Sources

Ray-tracing analyses have been conducted to search for any possible straylight source producing the in-field straylight contribution. Here we discuss two possible in-field straylight sources: rear filter back reflection and filter pinhole. None is however dominant, indicating that in-field straylight is probably dominated by wings of the point spread function (PSF), as for other similar EUV instruments (SOHO/EIT, STEREO/EUVI, TRACE) where significant power in their PSF wings has been observed (Defise, 1999, Auchère, 2000, DeForest *et al.*, 2009, Shearer *et al.* 2012).

#### 3.3.1. Rear Filter Back Reflection

In order to make EUV photons visible to SWAP's detector, the detector is coated with a scintillator material that absorbs EUV photons and re-emits visible photons. In fact, this coating re-emits visible photons in all directions, and it is possible that some of these re-emitted photons could be reflected back to the detector by the surfaces located in front of the detector. We estimated the level of such visible straylight using ray-tracing simulations.

One particular surface that could reflect these visible photons is the rear aluminum foil filter, located a few centimeters in front of the detector, as shown in Figure 6. Due to the filter inclination with respect to the optical axis, however, no specular reflection can reach directly the detector, as is shown on Figure 27a.

A cold cup, which is used to limit molecular contamination that might be trapped by the cooled detector, is mounted in front of the detector, as shown on Figure 6. Assuming 50 % of total photons are retro-emitted, a 1% black diffusive cold cup, and 80 % filter specular reflection, the ray-tracing model indicates that only 0.01 % of the in-the-field rays (*i.e.* solar signal) reach the detector after a specular reflection on the rear filter.

The straylight level due to specular reflection on rear filter is thus much lower that the observed background level.

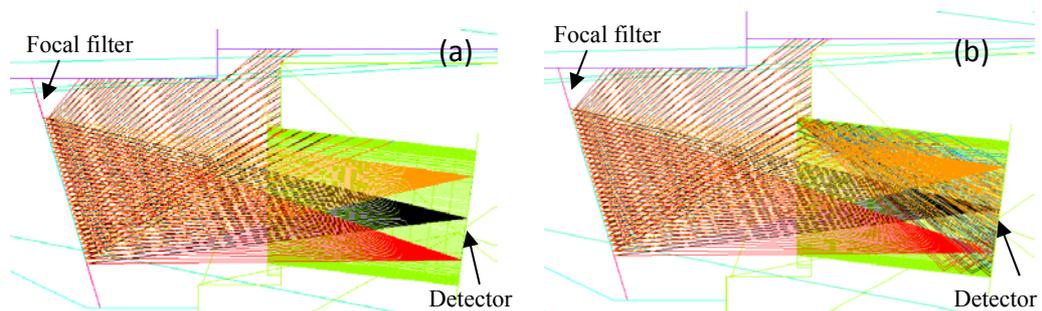

Figure 27: Rear filter specular reflection of retro-emitted photons by the detector scintillator coating (a) without and (b) with cold cup diffusion.

Furthermore the straylight resulting from reflection on the rear filter is not uniform, as shown by the ray-traced pattern (Figure 28) on the detector.

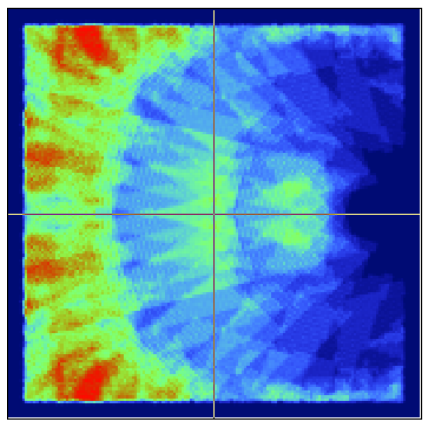

Figure 28: Rear filter straylight pattern on the SWAP detector array due to specular reflection of retro-emitted photons by the detector scintillator coating.

The straylight contribution from the diffusive reflection of retro-emitted photons at detector level, assuming 20 % filter scattering is only 0.0075 % of the in-field rays (*i.e.* solar signal).

The overall rear-filter straylight level is thus some orders of magnitude lower than the observed background level that appears in Figure 25.

### 3.3.2. Filter Pinholes

We also used the SWAP ray-tracing model to simulate a pinhole in the aluminum foil filter located near the instrument entrance, as shown in Figure 29, which would result in visible straylight inside the instrument cavity.

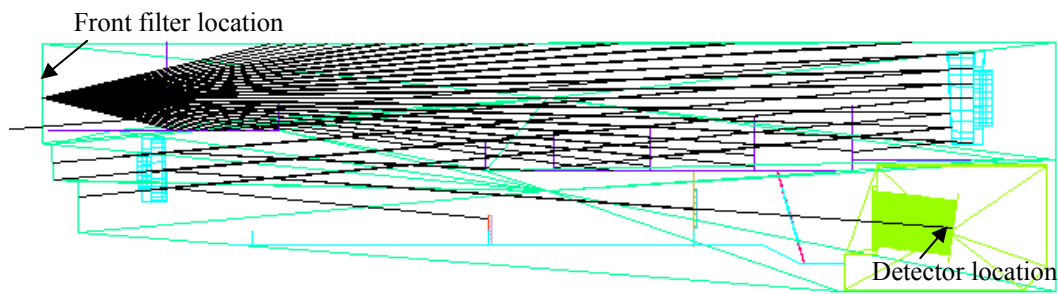

Figure 29: Simulation of a pinhole in the front filter (box is 565 x 150 x 125 mm).

For each pinhole, the resulting flux on the detector is 0.12 % of the incident flux. This value must be multiplied by the solid angle between the pinhole and the entrance pupil located a few centimeters in front of it ($2\pi/8$ steradian). The resulting straylight level on the detector due to a pinhole in the front filter would thus be 0.015 %, which is much lower than the observed background level of Figure 25. In addition, the rear filter provides a redundant protection for visible light. A pinhole in the front filter, without a pinhole in the back filter should thus not produce any visible straylight. It is possible that some parasitic reflections of the Sun within the instrument cavity produce straylight, but this could only happen during off-points (there is no reflection of the Sun in nominal pointing) or be due to scattering by particulate contamination on the two mirrors. In any case, such contributions are much lower than the observed background level of Figure 25.

## 4. Conclusions

The SWAP detector is the first scientific APS-CMOS used for scientific solar observation from space. On complement to preflight, ground-based calibration, in-flight measurements intended to characterize the detector performance and evolution have been performed regularly since launch. In particular, measurements to establish limits on detector noise and degradation have both improved calibrated data products and shown that CMOS-APS detectors can perform with little or no degradation over the long-term. Regular calibrations using LEDs have are also performed to monitor the performance of the instrument, showing a degradation lower than 2 % per year that is most probably related to LED ageing.

The in-flight detector calibration will be continued, including a thorough PSF analysis that should help to understand and remove additional straylight from SWAP images and improve far-field image contrast. Linearity analysis will also be continued, including in particular saturated images to investigate the saturation behavior as it nears full well. SWAP thus demonstrates the suitability of APS detector and of scintillator coating for scientific mission, as a precursor of the EUI instrument onboard the ESA *Solar Orbiter* mission.

Dedicated in-flight straylight calibrations have also been performed to evaluate the out-of-field straylight, which was found low enough for the regular dark and LED calibration sequences.

To confirm the out-of field straylight results, additional image sequences will be captured with off-point in the four directions and with more images per off-point angle to avoid detector-coating lag effects.

Two common possible in-field straylight contributions have also been analyzed to determine the source of a permanent background contribution within the images, but appeared to be much lower than the observed background. This straylight is therefore most probably the result of an enlarged point-spread-function due to scattering on mirrors that is not attenuated by the internal baffles. Deconvolution techniques will be applied to SWAP images to correct this artifact.

**Acknowledgements**. The SWAP instrument was developed by Centre Spatial de Liège (University of Liège) in collaboration with the Royal Observatory of Belgium. Support for calibration was provided by the Max Planck Institute for Solar System Research. Belgian activities are funded by the Belgian Federal Science Policy Office (BELSPO), through the ESA/PRODEX program for the payload instruments, and the ESA/GSTP program for the PROBA2 platform.